\documentclass[%
 reprint,
 amsmath,amssymb,
 prd,
]{revtex4-2}

\usepackage{graphicx}
\usepackage{dcolumn}
\usepackage{bm}

\usepackage{float}
\usepackage{color}

\usepackage{txfonts}
\usepackage{wrapfig}
\usepackage{multirow}
\usepackage{epsfig}
\usepackage{xcolor}
\usepackage{tcolorbox}
\usepackage{stackengine}
\usepackage{tabularx}
\usepackage{caption}
\usepackage{subcaption}
\usepackage{hyperref}
\usepackage{diagbox}
\usepackage{makecell}
\usepackage{nicematrix}
\newcommand{\apjs}{Astro. Phys. Journal Supliment}
\newcommand{\aj}{Astrophysical Journal}
\newcommand{\mnras}{MNRAS}
\newcommand{\aap}{Astronomy and Astrophysics}
\begin{document}


\title{Can the second time-derivative of the orbital frequency of binary pulsars\\ be used for testing general relativity?}


\author{Dhruv Pathak}
\email{dhruv.pathak@iucaa.in}
\author{Debarati Chatterjee}%
 \affiliation{%
Inter-University Centre for Astronomy and Astrophysics, Pune University Campus, Pune - 411007, India}%

\date{\today}

\begin{abstract}
With precision pulsar timing, measured values of a large set of pulsar parameters are obtainable. For some of those parameters, such as the time-derivatives of spin or orbital periods (in the case of binary pulsars), the measured values are not the intrinsic values of the parameters as they contain contributions from the dynamical effects. In the case of orbital period derivatives, the intrinsic values are essentially the general relativistic results. Pulsar timing solution also provides measurement of higher time-derivatives of orbital frequency for some pulsars. We specifically focus on the second time-derivative of the orbital frequency to explore its application in testing general relativity. In this work, we have provided a formalism to estimate the general relativistic contribution to the second derivative of the orbital frequency. We have calculated the dynamical effect contributions as well as the general relativistic contributions to the second time-derivative of the orbital period for real as well as synthetic pulsars. We find that the general relativistic contribution to the second time-derivative of the orbital period is negligibly small compared to the observed values of the real pulsars.
\end{abstract}

\maketitle


\section{Introduction}

Pulsar timing analysis provides measurements of pulsar parameters \cite{lk05}. In the case of binary pulsars, pulsar timing can provide measurements of both i) Keplerian parameters: orbital period ($P_b$), projected semi-major orbital axis ($a_p \sin i$, with $i$ being the angle of inclination of the orbital plane), orbital eccentricity (e), longitude of periastron ($\varpi$), epoch of periastron passage ($T_0$), position angle of the ascending node ($\Omega_{\rm asc}$) as well as ii) Post-Keplerian parameters: orbital period decay ($\dot{P}_{\rm b}$), advance of periastron ($\dot{\varpi}$), Einstein delay parameter ($\gamma$), Shapiro delay range ($r$), Shapiro delay shape ($s$) \cite{lk05}. The measured value of the pulsar parameters like the orbital (as well as spin) period derivatives are affected by the dynamics of the pulsar, that is, its velocity and acceleration. The gravitational potential of the Galaxy is a source of these accelerations, which is common to all pulsars \cite{pb18}. In order to estimate the intrinsic values of these parameters, the contribution of the dynamical effects to the measured values should be taken into account. In addition to the dynamical effects, there might be some contribution from other effects like mass loss, tidal effects, variations in the quadrupole moment, and so on \cite{lk05,lvt11,kb23}. We call these the residual effects. 

The intrinsic value of the orbital period decay parameter ($\dot{P}_{\rm b,int}$) is essentially given by the following expression which is a consequence of general relativity \cite{peters,mmgw,lk05}. 

\begin{align}
    \dot{P}_{\rm b, GR} = -\frac{192 \pi}{5}\frac{G^{5/3}}{c^5}\left(\frac{2\pi}{P_{\rm b}}\right)^{5/3}\frac{m_1 m_2}{(m_1 + m_2)^{1/3}}\frac{(1+\frac{73}{24}e^2+\frac{37}{96}e^4)}{(1-e^2)^{7/2}}
    \label{eq:pbdotgr}
\end{align}

Here, $P_{\rm b}$ is the orbital period of the binary pulsar system, $m_1$ is the mass of the pulsar, $m_2$ is the mass of the companion, $e$ is the eccentricity, $c$ is the speed of light, $G$ is the gravitational constant, $\dot{P}_{\rm b, GR}$ is the derivative of the orbital period due to emission of Gravitational radiation from the binary system. Subscript `GR' denotes that the parameters are the consequence of General relativity and the dot over a parameter denotes its time-derivative. 

$\dot{P}_{\rm b, GR}$ is known to be an essential parameter in the literature for the test of general relativity \cite{tay81,tw82,tay92,dt92,tay95,de96,sac98,ksm06,ks21} by comparing the measured value of $\dot{P}_{\rm b}$ with the calculated value of $\dot{P}_{\rm b, GR}$. Another pulsar parameter, whose measurements are available for some pulsars from pulsar timing analysis, is the orbital frequency's second derivative ($\ddot{f}_{\rm b, obs}$). 

As seen in the ATNF pulsar catalogue \cite{mhth05}, $\ddot{f}_{\rm b, obs}$ values are available for only eight pulsars- PSRs J0023+0923, J1048+2339, J1731-1847, J2339-0533, J0024-7204J, J0024-7204V, J0024-7204O, and J0024-7204W. Two additional pulsars are reported to have measurements of $\ddot{f}_{\rm b, obs}$ values- PSRs J1723-2837 \cite{cr13}, and J2051-0827 \cite{go16}. Among these pulsars, five are black widows- PSRs J0023+0923 \cite{abb18}, J0024-7204J \cite{fr17}, J0024-7204O \cite{fr17}, J1731-1847 \cite{nbb14}, and J2051-0827 \cite{go16} and four are redbacks- PSRs J0024-7204W \cite{ri16}, J1048+2339 \cite{drc16}, J1723-2837 \cite{cr13}, and J2339-0533 \cite{pc15}.

Both black widow and redback pulsars are binary millisecond pulsars with small values of the orbital period ($\sim$ 0.5 day), where the strong wind from the pulsar keeps on evaporating the companion. Black widows are identified by companions of mass of around 0.05 ${\rm M_{\odot}}$ while redbacks are identified by companions of mass between 0.1 and 0.5 ${\rm M_{\odot}}$ \cite{go16}. Higher derivatives of the orbital frequencies of these systems are affected by the pulsar wind and intra-binary matter. However, to understand the contributions of these, one will first need to subtract the external dynamical effects, e.g., the effect of the gravitational potential of the Galaxy, etc \cite{go16}. 

In this work, we focus on the second time-derivative of the orbital frequency to explore its application in testing general relativity. We provide a formalism to estimate the general relativistic contribution to the second derivative of the orbital frequency. We estimate the dynamical effect contributions as well as the general relativistic contributions to $\ddot{f}_{\rm b, obs}$ for real as well as synthetic pulsars. 

In section 2, we present the formalism to derive the analytical expressions for various contributing components of the second time-derivative of the orbital frequency. In section 3, we present the results and in section 4, we present the conclusions.

\section{Analytical expressions for various contributions to the second time-derivative of the orbital frequency}

The observed values of the first ($\dot{f}_{\rm b, obs}$) and second ($\dot{f}_{\rm b, obs}$) derivatives of the orbital frequency contain contributions from the general relativity ($\dot{f}_{\rm b, GR}$ and $\ddot{f}_{\rm b, GR}$), dynamical effects ($\dot{f}_{\rm b, dyn}$ and $\ddot{f}_{\rm b, dyn}$) and additional `residual' terms ($\dot{f}_{\rm b, x}$ and $\ddot{f}_{\rm b, x}$), as given in following expressions, 

\begin{align}
\dot{f}_{\rm b, obs} = \dot{f}_{\rm b, GR} + \dot{f}_{\rm b, dyn} + \dot{f}_{\rm b, x}~,
    \label{eq:fbdotobs}
\end{align}
and, on taking time-derivative on both sides of the above equation,
\begin{align}
\ddot{f}_{\rm b, obs} = \ddot{f}_{\rm b, GR} + \ddot{f}_{\rm b, dyn} + \ddot{f}_{\rm b, x}~.
    \label{eq:fbddotobs}
\end{align}

To estimate $\dot{f}_{\rm b, dyn}$ and $\ddot{f}_{\rm b, dyn}$, 
\begin{align}
    \dot{f}_{\rm b, dyn} = f_{\rm b}\,\left(\frac{\dot{f}}{f}\right)_{\rm dyn} ~,
    \label{eq:fdotdyn}
\end{align}

and 

\begin{align}
    \ddot{f}_{\rm b, dyn} = f_{\rm b}\,\left( \frac{\ddot{f}}{f}\right)_{\rm dyn}~,
    \label{eq:fddotdyn}
\end{align}

where the ratios $\left(\frac{\dot{f}}{f}\right)_{\rm dyn}$ and $\left( \frac{\ddot{f}}{f}\right)_{\rm dyn}$ are the dynamical effect terms and their expressions are given by,

\begin{equation}
\left(\frac{\dot{f}}{f}\right)_{\rm dyn} =  - \left[ \frac{ (\vec{a}_{\rm p} - \vec{a}_{\rm s}) \cdot \widehat{n}_{\rm sp}}{c} + \frac{1}{c} (\vec{v}_{\rm p} - \vec{v}_{\rm s}) \cdot \frac{d}{dt} (\widehat{n}_{\rm sp}) \right] ~,
\label{eq:fdotex}
\end{equation} and

\begin{align}
\left( \frac{\ddot{f}}{f}\right)_{\rm dyn} = &-\left[ \frac{\left(\dot{\vec{a}}_{\rm p}-\dot{\vec{a}}_{\rm s}\right)\cdot \widehat{n}_{\rm sp}}{c} + 2\frac{\left(\vec{a}_{\rm p} - \vec{a}_{\rm s}\right)\cdot\dot{\widehat{n}}_{\rm sp}}{c} + \frac{(\vec{v}_{\rm p} - \vec{v}_{\rm s})\cdot\ddot{\widehat{n}}_{\rm sp}}{c} \right. \nonumber \\ 
&\left.+2\left(\frac{\dot{f}}{f} \right)_{\rm dyn}\left(\frac{\vec{a}_{\rm s}\cdot \widehat{n}_{\rm sp}}{c} + \frac{\vec{v}_{\rm s}\cdot\dot{\widehat{n}}_{\rm sp}}{c} \right) -2\left(\frac{\dot{f}}{f}\right)_{\rm dyn}\left(\frac{\dot{f}}{f}\right)_{\rm obs}\right] ~.
\label{eq:fddotex1}
\end{align}  

In Eqs. (\ref{eq:fdotex}) and (\ref{eq:fddotex1}), $\vec{v}_{\rm p}$ is the velocity of the pulsar, $\vec{v}_{\rm s}$ is the velocity of the Sun, $\widehat{n}_{\rm sp}$ is the unit vector from the Sun to the pulsar and is taken to be the radial direction, $\vec{a}_{\rm p}$ is the acceleration of the pulsar, $\vec{a}_{\rm s}$ is the acceleration of the Sun, and $c$ is the speed of light. We use in our calculations the expressions of $\left(\frac{\dot{f}}{f}\right)_{\rm dyn}$ and $\left(\frac{\ddot{f}}{f}\right)_{\rm dyn}$ derived in terms of observable quantities as shown in \cite{pb18,pb21,dp21}. The observable parameters required for estimating $\left(\frac{\dot{f}}{f}\right)_{\rm dyn}$ are Galactic longitude ($l$), Galactic latitude ($b$), distance ($d$), proper motion along Galactic longitude ($\mu_l$), proper motion along Galactic latitude ($\mu_b$), orbital period ($P_{\rm b}$), and observed value of orbital period derivative ($\dot{P}_{\rm b, obs}$). For estimating $\left(\frac{\ddot{f}}{f}\right)_{\rm dyn}$, in addition to the above-mentioned observable parameters, the observed value of the second derivative of the orbital frequency ($\ddot{f}_{\rm b, obs}$) and radial component of the relative velocity between the pulsar and the Sun ($v_{r}$) are also required as observables \cite{pb18,pb21,dp21}.

In order to obtain the expression for $\dot{P}_{\rm b, GR}$, \cite{peters} calculates the average power loss due to gravitational radiation as,
\begin{align}
    \dot{E}_{\rm avg} =& \frac{32 G^4 \mu^2 M^3}{5 c^5 a^5 \left(1-e^2\right)^{7/2}}\left(1+\frac{73}{24}e^2 + \frac{37}{96}e^4\right)~.
    \label{eq:edotavg0}
\end{align}

This $\dot{E}_{\rm avg}$ is used to obtain $\dot{P}_{\rm b, GR}$ as given in Eq. (\ref{eq:pbdotgr}). On similar lines, we analytically derive the expression for $\ddot{P}_{\rm b, GR}$, which turns out be,

\begin{align}
    \ddot{P}_{\rm b,GR} &= 15 P_{\rm b} \frac{a^2}{G^2 m_1^2 m_2^2} \dot{E}_{\rm avg}^2 
    \label{eq:nomlosspbddotgr0}
\end{align}

See Appendix \ref{app1} for details of the analytical derivation.

To estimate $\ddot{f}_{\rm b, GR}$ from period domain quantities $\ddot{P}_{\rm b, GR}$ and $\dot{P}_{\rm b, GR}$, we use, 
\begin{align}
\ddot{f}_{\rm b, GR} = -\frac{\ddot{P}_{\rm b, GR}}{P_{\rm b}^2} + 2\frac{\dot{P}_{\rm b, GR}^2}{P_{\rm b}^3}~.
    \label{eq:fbddotgr}
\end{align}

\section{Results}
\label{sec:res}
\subsection{Estimating $\ddot{f}_{\rm b, dyn}$ and $\ddot{f}_{\rm b, GR}$ values for real pulsars}
\label{sec:realpsrs}
As per the ATNF pulsar catalogue, there are 5 pulsars (PSRs J0348+0432, J0740+6620, J1738+0333, J2045+3633, and J2339-0533) with all the required parameters available for the calculation of $\ddot{f}_{\rm b,dyn}$ and $\ddot{f}_{\rm b,GR}$.  Additionally, \cite{ks21} provides all the parameters of the double pulsar PSR J0737-3039A required for the calculation of $\ddot{f}_{\rm b,dyn}$ and $\ddot{f}_{\rm b,GR}$ values. 

We calculate $\dot{f}_{\rm b,GR}$ using the expression for $\dot{P}_{\rm b,GR}$ given in Eq. (\ref{eq:pbdotgr}). We use a Python package, `GalDynPsr'\footnote{https://github.com/pathakdhruv/GalDynPsr} \cite{pb18} for estimating $\dot{f}_{\rm b,dyn}$ values following Eqs. (\ref{eq:fdotdyn}) and (\ref{eq:fdotex}). We specifically use Model-Lb of GalDynPsr for calculating $\left(\frac{\dot{f}}{f}\right)_{\rm dyn}$ and consequently $\dot{f}_{\rm b,dyn}$. Using the calculated values of $\dot{f}_{\rm b,GR}$ and $\dot{f}_{\rm b,dyn}$ as well as the available $\dot{f}_{\rm b,obs}$ value, we obtain the residual orbital frequency derivative value ($\dot{f}_{\rm b,x}$) using Eq. (\ref{eq:fbdotobs}). For the second derivative values, we use Eq. (\ref{eq:fbddotgr}) to calculate $\ddot{f}_{\rm b,GR}$. For calculating $\ddot{f}_{\rm b,dyn}$, we use Eqs. (\ref{eq:fddotdyn}) and (\ref{eq:fddotex1}) following the formalism described in \cite{pb18,pb21,dp21}. 

We use the values of the observable parameters ($l$, $b$, $d$, $\mu_l$, $\mu_b$, $P_{\rm b}$, $\dot{P}_{\rm b, obs}$, $\ddot{f}_{\rm b, obs}$) given in the ATNF pulsar catalogue for PSRs J0348+0432, J0740+6620, J1738+0333, J2045+3633, and J2339-0533 and in \cite{ks21} for PSR J0737-3039A. We take a nominal value of 50 km/s for $v_{r}$ as taken in \cite{liu19}. The values of $\dot{f}_{\rm b,dyn}$, $\dot{f}_{\rm b,GR}$, $\dot{f}_{\rm b,x}$, $\ddot{f}_{\rm b,dyn}$, and $\ddot{f}_{\rm b,GR}$ that we obtain for these six pulsars are displayed in Table \ref{tb:fddotgrreal}.  We can see that the maximum magnitude of $\ddot{f}_{\rm b,GR}$ is obtained for the double pulsar PSR J0737-3039A. We also see that $\ddot{f}_{\rm b,dyn}$ values are comparable to $\ddot{f}_{\rm b,GR}$ values for PSRs J0348+0432, J0737-3039A, and J2339-0533, whereas, for PSRs J0740+6620, J1738+0333, and J2045+3633 $\ddot{f}_{\rm b,GR}$  values are negligible compared to the respective $\ddot{f}_{\rm b,dyn}$ values. However, the values of both $\ddot{f}_{\rm b,GR}$ and $\ddot{f}_{\rm b,dyn}$ obtained in Table \ref{tb:fddotgrreal} are negligible compared to the range of $\ddot{f}_{\rm b, obs}$ values available in ATNF pulsar catalogue ($O(10^{-29})~s^{-3} <|\ddot{f}_{\rm b, obs}|< O(10^{-25}) ~s^{-3}$). We speculate that the major contribution to $\ddot{f}_{\rm b, obs}$ might be coming from the residual term. 

We also see that the magnitude of $\dot{f}_{\rm b,GR}$ values is much higher than $\dot{f}_{\rm b,x}$ values for PSRs J0348+0432, J0737-3039A, and J1738+0333. However, for PSR J0740+6620 both $\dot{f}_{\rm b,GR}$ as well as $\dot{f}_{\rm b,x}$ values have the same order of magnitude and for PSRs J2045+3633 and J2339-0533 the magnitude of $\dot{f}_{\rm b,GR}$ values is negligibly smaller than $\dot{f}_{\rm b,x}$ values indicating a significant contribution to orbital frequency derivative from sources other than gravitational radiation emission. These additional contributions can be due to mass loss, tidal effects, variations in the quadrupole moment, etc \cite{lk05,lvt11,kb23}.

\begin{table*}[h]
\begin{center}
\caption{Parameters for real pulsars with all the required parameters available for the calculation of $\ddot{f}_{\rm b,dyn}$ and $\ddot{f}_{\rm b,GR}$. Here, $f_{\rm b}$ is the orbital frequency, $\dot{f}_{\rm b,obs}$ is observed orbital frequency derivative,  $\ddot{f}_{\rm b,obs}$ is observed orbital frequency second derivative, $\dot{f}_{\rm b,dyn}$ is the dynamical effect contribution to the observed orbital frequency derivative, $\dot{f}_{\rm b,GR}$ is the general relativistic contribution to the observed orbital frequency derivative, $\dot{f}_{\rm b,x}$ is the residual orbital frequency derivative, $\ddot{f}_{\rm b,dyn}$ is the dynamical effect contribution to the orbital frequency second derivative, and $\ddot{f}_{\rm b,GR}$ is the general relativistic contribution to the orbital frequency second derivative.} 
\hspace*{-1cm}\begin{tabular}{|@{\hskip1.pt}l@{\hskip1.pt}|@{\hskip1.pt}l@{\hskip1.pt}|@{\hskip1.pt}l@{\hskip1.pt}|@{\hskip1.pt}l@{\hskip1.pt}|@{\hskip1.pt}l@{\hskip1.pt}|@{\hskip1.pt}l@{\hskip1.pt}|@{\hskip1.pt}l@{\hskip1.pt}|}
 \hline 
     \diagbox{Parameters}{PSR} & J0348+0432 & J0737-3039A & J0740+6620 & J1738+0333 & J2045+3633 & J2339-0533  \\  
\hline
$f_{\rm b}$ (Hz)  & 1.13002$\times10^{-4}$  &  1.13192$\times10^{-4}$  &  2.4280$\times10^{-6}$  &  3.26223$\times10^{-5}$  &  3.58354$\times10^{-7}$  &  5.99387$\times10^{-5}$ \\   
\hline
\addstackgap{$\dot{f}_{\rm b,obs}$ ($\rm s^{-2}$)} & 3.5$\times10^{-21}$ &  1.60$\times10^{-20}$  &  -7.1$\times10^{-24}$  &  1.8$\times10^{-23}$  &  -4.5$\times10^{-25}$  &  5.97$\times10^{-19}$ \\     
\hline   
& & & & & & \\
\addstackgap{$\dot{f}_{\rm b,dyn}$ ($\rm s^{-2}$)} & -2.2$\times10^{-23}$  &  1.6$\times10^{-24}$  &  -7.1$\times10^{-24}$  &  -1.2$\times10^{-23}$  &  1.3$\times10^{-25}$  &  4.0$\times10^{-25}$ \\   
\hline 
\addstackgap{$\dot{f}_{\rm b,GR}$ ($\rm s^{-2}$)} & 3.3$\times10^{-21}$  &  1.6$\times10^{-20}$  &  3.9$\times10^{-27}$  &  2.9$\times10^{-23}$  &  7.3$\times10^{-30}$  &  5.5$\times10^{-22}$\\  
\hline 
\addstackgap{$\dot{f}_{\rm b,x}$ ($\rm s^{-2}$)} & 2.1$\times10^{-22}$  &  -4.9$\times10^{-25}$  &  9.7$\times10^{-27}$  &  8.0$\times10^{-25}$  &  -5.8$\times10^{-25}$  &  6.0$\times10^{-19}$ \\ 
\hline 
\addstackgap{$\ddot{f}_{\rm b,dyn}$ ($\rm s^{-3}$)} & -1.8$\times10^{-38}$  &  1.8$\times10^{-38}$  &  2.6$\times10^{-38}$  &  5.3$\times10^{-38}$  &  -3.1$\times10^{-41}$  &  8.1$\times10^{-38}$ \\ 
\hline 
\addstackgap{$\ddot{f}_{\rm b,GR}$ ($\rm s^{-3}$)} & 3.2$\times10^{-38}$  &  7.5$\times10^{-37}$  &  2.1$\times10^{-48}$  &  8.8$\times10^{-42}$  &  4.9$\times10^{-53}$  &  1.7$\times10^{-39}$ \\ 
\hline
\end{tabular}
\label{tb:fddotgrreal}
\end{center}
\end{table*}

\vskip1cm 
\subsection{Estimating residual $\ddot{f}_{\rm b}$ value for PSR J2339-0533} 
\label{sec:J2339}
As per the ATNF pulsar catalogue, we found that PSR J2339-0533 is the only pulsar with a measurement of $\ddot{f}_{\rm b, obs}$ that also has all the parameter values available required for the calculation of $\ddot{f}_{\rm b, GR}$ and $\ddot{f}_{\rm b, dyn}$ terms. We calculate $\ddot{f}_{\rm b, dyn}$ from Eq. (\ref{eq:fdotex}) and Eq. (2.37) of \cite{dp21}, and $\ddot{f}_{\rm b, GR}$ from Eqs. (\ref{eq:nomlosspbddotgr0}) and (\ref{eq:fbddotgr}). We then estimate the residual $\ddot{f}_{\rm b}$ value, i.e., $\ddot{f}_{\rm b, x}$ by subtracting from $\ddot{f}_{\rm b, obs}$ the sum of $\ddot{f}_{\rm b, GR}$ and $\ddot{f}_{\rm b, dyn}$ (eq. \ref{eq:fbddotobs}).

 Hence, given $\ddot{f}_{\rm b, obs}=1.77\times10^{-26} ~s^{-3}$, we compute $\ddot{f}_{\rm b, GR} = 1.677\times10^{-39} ~s^{-3}$ and $\ddot{f}_{\rm b, dyn} = 8.111\times10^{-38} ~s^{-3}$, and consequently obtain $\ddot{f}_{\rm b, x} = 1.77\times10^{-26} ~s^{-3}$. In the above calculations, we took a nominal value of $v_{r}$ as 50 km/s. With change in $v_{r}$, we do observe change in $\ddot{f}_{\rm b, dyn}$ value but there is no significant change in the $\ddot{f}_{\rm b, x}$ value. For $v_{r}=-200$ km/s, $\ddot{f}_{\rm b, dyn}=-2.125\times10^{-37}~s^{-3}$, $\ddot{f}_{\rm b, x}=1.77\times10^{-26}~s^{-3}$, and for $v_{r}=200$ km/s, $\ddot{f}_{\rm b, dyn}=2.573\times10^{-37}~s^{-3}$ and $\ddot{f}_{\rm b, x}=1.77\times10^{-26}~s^{-3}$.

We can see that both $\ddot{f}_{\rm b, GR}$ and $\ddot{f}_{\rm b, dyn}$ values are negligible compared to $\ddot{f}_{\rm b, x}$. Consequently, the main contribution to $\ddot{f}_{\rm b, obs}$ values is coming from the residual term $\ddot{f}_{\rm b, x}$, which includes mass loss effects, tidal effects, or time-derivative of quadrupole moment variation. This also further consolidates our speculation in Section \ref{sec:realpsrs}. As far as real pulsars are concerned, we can say that comparing values of $\ddot{f}_{\rm b, GR}$ and $\ddot{f}_{\rm b, obs}$ is not a viable test of general relativity because $\ddot{f}_{\rm b, GR}$ is negligible compared to the observed value and the main contribution to the observed value is from the residual terms.

Note that, in this work, we are concerned only with the order of magnitude comparison among the terms contributing to $\ddot{f}_{\rm b, obs}$. Hence, we have not reported any error terms in Tables \ref{tb:fddotgrreal} and \ref{tb:fddotgrsim}. For a discussion on error calculation of $\ddot{f}_{\rm b, GR}$ value of real pulsars using standard error propagation technique, refer to the appendix \ref{app:err}.

\subsection{Exploring $\ddot{f}_{\rm b, GR}$ values for simulated pulsars}
\label{sec:simpsr}
As mentioned previously, as per the ATNF pulsar catalogue, measured values of the second derivative of the orbital frequency ($\ddot{f}_{\rm b, obs}$) are available only for 8 pulsars. We wished to explore the range $\ddot{f}_{\rm b, GR}$ values can take for a set of simulated pulsars and whether this component of $\ddot{f}_{\rm b}$ can be large enough to be a measurable quantity.

For the simulated pulsars, we use the simulated values of the observables ($l$, $b$, $d$, $\mu_l$, $\mu_b$, $f_{\rm b}$, $\dot{f}_{\rm b, obs}$) such that they follow the same probability density distribution as the values for real pulsars available in the ATNF pulsar catalogue (excluding the ones in the globular clusters, the Large Magellanic Cloud, and the Small Magellanic Cloud as in these cases, there will be extra dynamical effects due to the local gravitational potentials).

To ensure that the parameter for the synthetic and the real pulsars follow the same distribution, we first fit an Empirical Cumulative Distribution Function (ECDF) to the distribution followed by the real pulsar values and then used the inverse CDF technique to generate the intended number of synthetic values following the same ECDF. This technique is discussed in detail in \cite{pb21,dp21}. We generate 10000 values of the observable parameters using this technique. For $v_r$, we took a uniform distribution between -200 and 200 km/s, the same as the range used by \cite{liu18}.

\begin{figure*}[h]
\centering
\includegraphics[width=0.8\textwidth]{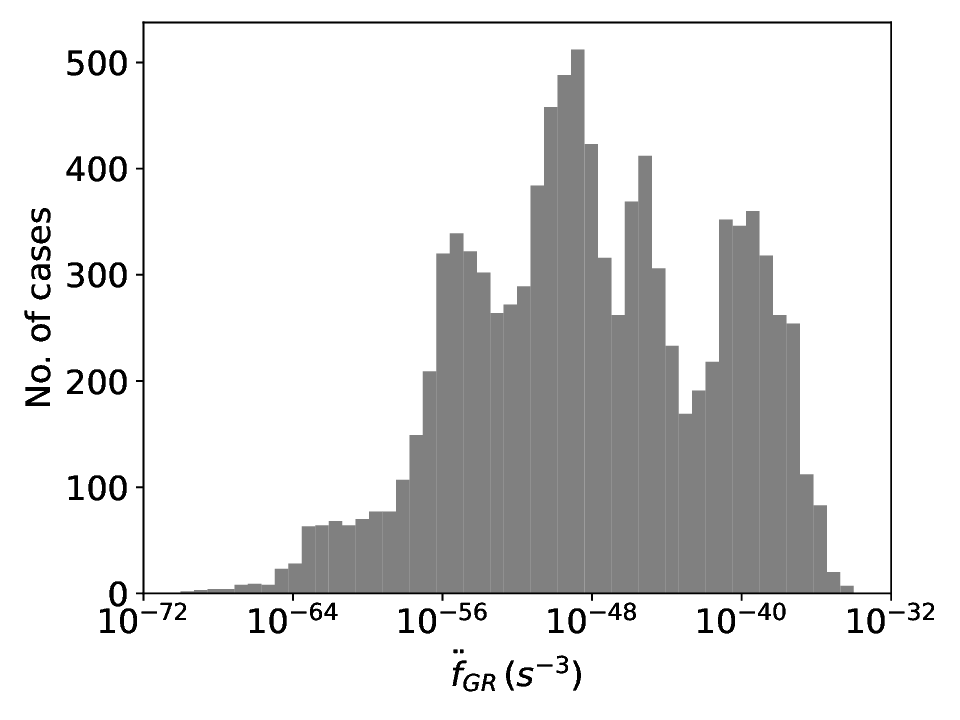}
\caption{Histogram plot showing the number of simulated pulsars for different ranges of  $\ddot{f}_{\rm b, GR}$ values.}
\label{fig:hist}
\end{figure*}

We calculate the values of $\dot{f}_{\rm b,dyn}$, $\dot{f}_{\rm b,GR}$, $\dot{f}_{\rm b,x}$, $\ddot{f}_{\rm b,dyn}$, and $\ddot{f}_{\rm b,GR}$ as discussed in section \ref{sec:realpsrs}. In Table \ref{tb:fddotgrsim}, the observables as well as the calculated parameters for the top six simulated pulsars out of the 10000 are displayed for which $\ddot{f}_{\rm b,GR}$ has the highest magnitude. Fig. \ref{fig:hist}
shows the histogram plot for the number of cases versus the $\ddot{f}_{\rm b,GR}$ values of the simulated pulsars. 

We find that the maximum order of magnitude of $\ddot{f}_{\rm b,GR}$, obtained for the simulated pulsars is $10^{-35} ~s^{-3}$ and there are 13 cases with this order of magnitude. The minimum order of magnitude obtained is $10^{-71} ~s^{-3}$ (one case). Furthermore, for the absolute value of $\ddot{f}_{\rm b,dyn}$, the maximum order of magnitude is $10^{-33} ~s^{-3}$ (one case) and the minimum order of magnitude obtained is $10^{-43} ~s^{-3}$ (15 cases). We know from ATNF pulsar catalogue that $\ddot{f}_{\rm b,obs}$ values of the real pulsars vary between the orders of magnitude $10^{-29}~s^{-3}$ and $10^{-25}~s^{-3}$, and hence, both $\ddot{f}_{\rm b,GR}$ and $\ddot{f}_{\rm b,dyn}$ are negligible compared to the range of observed values available in ATNF pulsar catalogue. Even for the simulated pulsars, we did not come across any case where $\ddot{f}_{\rm b,GR}$ is comparable to available observed values. This result also indicates that comparing $\ddot{f}_{\rm b,GR}$ and $\ddot{f}_{\rm b,obs}$ is not a viable test of general relativity.

\begin{table*}
\caption{Parameters for top six simulated pulsars with maximum $\ddot{f}_{\rm b,GR}$ values. Here, $l$ is the Galactic longitude, $b$ is the Galactic latitude, $d$ is the distance between the pulsar and the solar system barycentre, $d_{\rm GC}$ is the distance between the pulsar and the Galactic centre, $\mu_{l}$ is proper motion in $l$, $\mu_{b}$ is proper motion in $b$, $v_{r}$ is the radial velocity, $f_{\rm b}$ is the orbital frequency, $\dot{f}_{\rm b,obs}$ is observed orbital frequency derivative, $\dot{f}_{\rm b,dyn}$ is the dynamical effect contribution to the observed orbital frequency derivative, $\dot{f}_{\rm b,GR}$ is the general relativistic contribution to the observed orbital frequency derivative, $\dot{f}_{\rm b,x}$ is the residual orbital frequency derivative, $\ddot{f}_{\rm b,dyn}$ is the dynamical effect contribution to the orbital frequency second derivative, $\ddot{f}_{\rm b,GR}$ is the general relativistic contribution to the orbital frequency second derivative, $\ddot{f}_{\rm b,x}$ is the residual orbital frequency second derivative. Displaying the results till the fifth decimal place.}
\begin{tabular}{|@{\hskip1.pt}l@{\hskip1.pt}|@{\hskip1.pt}l@{\hskip1.pt}|@{\hskip1.pt}l@{\hskip1.pt}|@{\hskip1.pt}l@{\hskip1.pt}|@{\hskip1.pt}l@{\hskip1.pt}|@{\hskip1.pt}l@{\hskip1.pt}|@{\hskip1.pt}l@{\hskip1.pt}|}
 \hline 
 Parameters & Pulsar1 & Pulsar2 & Pulsar3 & Pulsar4 & Pulsar5 & Pulsar6 \\ 
[1em]\hline 
$l$ (deg)  & 299.088 &  270.096 & 194.321  & 353.69 & 279.652 &  245.207  \\ 
\hline
$b$ (deg) & -38.3158 &  10.7324 & -9.13854  & 8.419 & -2.061 &    -42.2317   \\
\hline
$d$ (kpc) & 1.80998  &  2.56655  &  0.710394  & 2.33001 & 7.3273 &    1.18352  \\ 
\hline
$d_{\rm GC}$ (kpc) & 7.49866  &  8.39758  &  8.68205  & 5.72486 & 7.91028 &    8.44277 \\ 
\hline
$\mu_{l}$ (mas/yr)  & 5.58780  &  7.76355  &  1.35880  & -13.0835 & -1.08543 &    1.30168 \\ 
\hline
$\mu_{b}$ (mas/yr)  & -1.37472  &  7.91087  &  -3.14688  & -3.41706  & 14.0564 &   -2.77816  \\     
\hline
$v_{r}$ (km/s) & -128.855 &  124.821  &  -85.3558  &  -107.369  &  -176.950  &  -152.756 \\ 
\hline
$m_1$ ($M_{\odot}$) & 1.66332  &  1.54198  &  2.13781 &  2.22080  &  2.10393  &  1.30986  \\ 
\hline
$m_2$ ($M_{\odot}$) & 1.54190  &  1.28758  &  0.956221 &  1.22143  &  1.21655 &   0.943866 \\ 
\hline
$e$  & 1.69748$\times10^{-4}$  &  3.08214$\times10^{-2}$  &  2.57052$\times10^{-6}$  &   5.03410$\times10^{-5}$  &  5.94201$\times10^{-7}$  &    5.01117$\times10^{-5}$  \\ 
\hline
$f_{\rm b}$ (Hz)  & 2.22136$\times10^{-4}$  &  2.17831$\times10^{-4}$ &  2.15420$\times10^{-4}$  &  1.75298$\times10^{-4}$   &  1.77326$\times10^{-4}$   &    2.10511$\times10^{-4}$  \\   
\hline
\addstackgap{$\dot{f}_{\rm b,obs}$ ($\rm s^{-2}$)} & -2.34009$\times10^{-24}$ &  -2.77399$\times10^{-25}$  &  3.10893$\times10^{-19}$  &  -1.69264$\times10^{-22}$  &  -2.43905$\times10^{-21}$  &    1.73706$\times10^{-19}$ \\  
\hline  
\addstackgap{$\dot{f}_{\rm b,dyn}$ ($\rm s^{-2}$)} & 1.94967$\times10^{-23}$  &  -1.22066$\times10^{-22}$  &  -1.48764$\times10^{-23}$  &  -2.21310$\times10^{-22}$  &  -5.22432$\times10^{-23}$  &    2.17377$\times10^{-23}$ \\   
\hline 
\addstackgap{$\dot{f}_{\rm b,GR}$ ($\rm s^{-2}$)} & 2.57367$\times10^{-19}$  &  1.94514$\times10^{-19}$ &  1.85470$\times10^{-19}$  &  1.11553$\times10^{-19}$  &  1.11120$\times10^{-19}$  &  1.14563$\times10^{-19}$ \\  
\hline 
\addstackgap{$\dot{f}_{\rm b,x}$ ($\rm s^{-2}$)} & -2.57388$\times10^{-19}$   &  -1.94393$\times10^{-19}$   &  1.25438$\times10^{-19}$   &  -1.11501$\times10^{-19}$   &  -1.13507$\times10^{-19}$   &  5.91211$\times10^{-20}$  \\ 
\hline  
\addstackgap{$\ddot{f}_{\rm b,dyn}$ ($\rm s^{-3}$)} & -3.10210$\times10^{-37}$   &  1.01596$\times10^{-36}$  &  -8.11456$\times10^{-38}$  &  -8.28675$\times10^{-37}$  &  -1.57815$\times10^{-36}$ & -2.91486$\times10^{-38}$ \\ 
\hline
\addstackgap{$\ddot{f}_{\rm b,GR}$ ($\rm s^{-3}$)} & 9.93950$\times10^{-35}$ &  5.78979$\times10^{-35}$  &  5.32282$\times10^{-35}$  &  2.36629$\times10^{-35}$  &  2.32111$\times10^{-35}$  &  2.07825$\times10^{-35}$  \\ 
\hline 
\end{tabular}
\label{tb:fddotgrsim}
\end{table*}

\clearpage

\section{Conclusion}

In this work, we have explored the second derivative of the orbital frequency of binary pulsars. We give the analytical expression for $\ddot{P}_{\rm b, GR}$ following the formalism for deriving the expression for $\dot{P}_{\rm b, GR}$ as given in \cite{peters,mmgw}.

As discussed in section \ref{sec:realpsrs}, there are six pulsars (PSRs J0348+0432, J0740+6620, J1738+0333, J2045+3633, J2339-0533, and J0737-3039A) with all the required parameters available for the calculation of $\ddot{f}_{\rm b,dyn}$ and $\ddot{f}_{\rm b,GR}$. We estimate $\ddot{f}_{\rm b,dyn}$ and $\ddot{f}_{\rm b,GR}$ values for these pulsars and find that $\ddot{f}_{\rm b,dyn}$ values are comparable to $\ddot{f}_{\rm b,GR}$ values for all the cases. Out of all these pulsars, the maximum absolute value of $\ddot{f}_{\rm b,GR}$ is obtained for the double pulsar PSR 0737-3039A. We also calculate the value of the residual term $\ddot{f}_{\rm b,x}$ for PSR 2339-0533 as this the only pulsar with a measurement of $\ddot{f}_{\rm b, obs}$ that also has all the parameter values available required for the calculation of $\ddot{f}_{\rm b, GR}$ and $\ddot{f}_{\rm b, dyn}$ terms. We obtain $\ddot{f}_{\rm b, x} = 1.77\times10^{-26} ~s^{-3}$ with negligible contribution from $\ddot{f}_{\rm b, GR}$ and $\ddot{f}_{\rm b,dyn}$ terms. We conclude that the main contribution to the observed value is coming from the residual term, which includes mass loss effects, tidal effects, or time-derivative of quadrupole moment variation.

We also compare the contributions from $\ddot{f}_{\rm b,dyn}$,  $\ddot{f}_{\rm b,GR}$, and $\ddot{f}_{\rm b,x}$ to $\ddot{f}_{\rm b,obs}$ for 10000 simulated pulsars. We find that $\ddot{f}_{\rm b,GR}$ values vary between the orders of magnitude $10^{-71}~s^{-3}$ and $10^{-35}~s^{-3}$ whereas absolute values of $\ddot{f}_{\rm b,dyn}$ vary between the orders of magnitude $10^{-43}~s^{-3}$ and $10^{-33}~s^{-3}$. As the absolute values of $\ddot{f}_{\rm b,obs}$ of the real pulsars vary between the orders of magnitude $10^{-29}~s^{-3}$ and $10^{-25}~s^{-3}$, both $\ddot{f}_{\rm b,dyn}$ and  $\ddot{f}_{\rm b,GR}$ terms are negligible compared to the range of $\ddot{f}_{\rm b,obs}$ values observed for real pulsars. We speculate the dominant contribution to the $\ddot{f}_{\rm b,obs}$ values comes from the residual terms and hence comparing the $\ddot{f}_{\rm b,GR}$ and $\ddot{f}_{\rm b,obs}$ values is not a viable test of general relativity. Only when we have realistic estimates of $\ddot{f}_{\rm b,x}$, we can discuss having a potential test of general relativity. If we have realistic estimates of the residual terms $\ddot{f}_{\rm b,x}$, then we can add to it the dynamical effect contribution $\ddot{f}_{\rm b,dyn}$ and we can then subtract this sum from the measured values $\ddot{f}_{\rm b,obs}$. The result should give the general relativistic contribution, which, in principle, should be equal to $\ddot{f}_{\rm b,GR}$ value calculated using the formalism described in this work. However, for estimating $\ddot{f}_{\rm b,x}$ and also $\dot{f}_{\rm b,x}$ we should calculate the individual contributions due to effects of mass loss, tidal effects, variations in quadrupole moment and their time-derivatives. With the advent of the Square Kilometre Array (SKA)\footnote{https://www.skao.int/en} it is expected that around 14000 normal pulsars and 6000 millisecond pulsars will be discovered \cite{sk09}. It may be possible that with such an increase in the pulsars, we will have more measurements of $\ddot{f}_{\rm b,obs}$ values and the range of orders of magnitude of $\ddot{f}_{\rm b,obs}$ also increases. Even then it seems unlikely that $\ddot{f}_{\rm b,GR}$ will become the dominant contributor to the observed values given the maximum order of magnitude we obtained for the simulated pulsars was as low as $10^{-33}~s^{-3}$.

\begin{acknowledgments}
This work makes use of \texttt{GNU Scientific Library}, \texttt{NumPy}~\cite{Numpy} , \texttt{Matplotlib}~\cite{Matplotlib}, and \texttt{galpy}~\cite{bovy15} software packages.
\end{acknowledgments}

\appendix
\section{Derivation of $\ddot{P}_{\rm b,GR}$}
\label{app1}

To obtain the expression of $\dot{P}_{\rm b, GR}$, we begin by taking the binary orbit in the x-y plane. We consider the binary components to have masses $m_1$ and $m_2$. We study the binary system as the effective one-body system of a reduced mass ($\mu$) orbiting in an elliptical orbit of radial distance $r$ around a source of gravitational potential of effective mass $M=m_1+m_2$, situated at one of the foci of the ellipse. The semi-major axis is denoted by $a$, the eccentricity by $e$, and the angle the radius vector makes with the semi-major axis is denoted by $\theta$. $r$ is given by the relation,
\begin{align}
    r = \frac{a\,(1 - e^2)}{1 + e\, \cos \theta}~,
\end{align}
and $\dot{\theta}$ is given by,
\begin{align}
    r^2 \dot{\theta} = \left\{GM a(1-e^2) \right\}^{1/2} ~.
    \label{eq:thetadot}
\end{align}
Hence, $\dot{r}$ becomes,
\begin{align}
    \dot{r} = \frac{a\,(1 - e^2) e \sin \theta \,\dot{\theta}}{(1 + e\, \cos \theta)^2}~.
\end{align}

The components of the moment of inertia ($I$) are given by the following expressions,

\begin{align}
    I_{xx} &= \mu r^2 \cos^2 \theta ,\\
    I_{yy} &= \mu r^2 \sin^2 \theta ,\\
    I_{xx} + I_{yy} &=  \mu r^2,\\
    I_{xy} &= \mu r^2 \sin \theta \, \cos \theta ,
\end{align}
where $\mu = \frac{m_1\, m_2}{m_1 + m_2}$.

Taking the third time-derivative of the moment of inertia components, we get,
\begin{align}
    \dddot{I}_{xx} &= \left(\frac{4G^3 \mu^2 M^3}{a^5 (1-e^2)^5}\right)^{1/2} \,(1+e \cos \theta)^2\, \left(2\sin 2\theta + 3e\sin \theta \cos^2 \theta\right), \\
    \dddot{I}_{yy} &= \left(\frac{4G^3 \mu^2 M^3}{a^5 (1-e^2)^5}\right)^{1/2} (1+e \cos \theta)^2 \left(-2\sin 2\theta - 3e\sin \theta \cos^2 \theta -e\sin \theta\right),\\
    \dddot{I}_{xy} &= \left(\frac{4G^3 \mu^2 M^3}{a^5 (1-e^2)^5}\right)^{1/2} \,(1+e \cos \theta)^2\, \left(-2\cos 2\theta - 3e \cos^3 \theta + e\cos \theta\right)~.
\end{align}
Taking the fourth time-derivative of the moment of inertia components, we get,
\begin{align}
    \ddddot{I}_{xx} &= \frac{2G^2 \mu M^2}{a^3 (1-e^2)^3} \,(1+e \cos \theta)^3\,\frac{1}{8} \left(32\cos2\theta  \right. \nonumber \\
    &\left.+ (6 \cos\theta + 50\cos3\theta)\,e - (3 - 12 \cos2\theta - 15\cos4\theta)\,e^2\right), \\
    \ddddot{I}_{yy} &= -\frac{2G^2 \mu M^2}{a^3 (1-e^2)^3} \,(1+e \cos \theta)^3\,\frac{1}{8} \left(32\cos2\theta \right. \nonumber \\
    &\left.+ (14 \cos\theta + 50\cos3\theta)\,e  - (7 - 24\cos2\theta - 15\cos4\theta)\,e^2\right),\\
    \ddddot{I}_{xy} &= \frac{2G^2 \mu M^2}{a^3 (1-e^2)^3} \,(1+e \cos \theta)^3\,\frac{\sin\theta}{4} \left(32\cos\theta  \right. \nonumber \\
    &\left.+ (30 + 50\cos2\theta)\,e + (33\cos\theta + 15\cos3\theta)\,e^2\right)~.
\end{align}

Following the derivation of $\dot{P}_{\rm b,GR}$ given in \cite{peters,mmgw}, we need to first calculate the average power loss $\dot{E}_{\rm avg}$. Instantaneous power loss is given by,

\begin{align}
    \dot{E} = \frac{G}{5c^5} \left[ \dddot{I}_{xx}^2 + \dddot{I}_{yy}^2 + 2 \dddot{I}_{xy}^2 - \frac{1}{3}\left(\dddot{I}_{xx} + \dddot{I}_{yy}\right)^2\right]~.
\end{align}

\vskip0.5cm
The average power loss is given by,
\begin{align}
    \dot{E}_{\rm avg} = \frac{1}{P_{\rm b}} \int_{0}^{P_{\rm b}} \dot{E} \,dt
\end{align}
This gives,
\begin{align}
    \dot{E}_{\rm avg} =& \frac{32 G^4 \mu^2 M^3}{5 c^5 a^5 \left(1-e^2\right)^{7/2}}\left(1+\frac{73}{24}e^2 + \frac{37}{96}e^4\right)~.
    \label{eq:edotavg}
\end{align}

From Kepler's third law and taking the average on both sides, we get,

\begin{align}
    \dot{P}_{\rm b,GR} =  \frac{3}{2} \frac{\dot{a}_{\rm avg}}{a} P_{\rm b} ~.
    \label{eq:Pbarel}
\end{align}

Now, we know,
\begin{align}
    a = \frac{G m_1 m_2}{2 E} = \frac{G \mu M}{2 E}
    \label{eq:aErel}
\end{align}

Taking time-derivative of Eq. (\ref{eq:aErel}), we get,
\begin{align}
    \dot{a} &= - \frac{G m_1 m_2}{2 E^2} \dot{E} \\
    \implies \dot{a} &= - \frac{2 a^2}{G m_1 m_2} \dot{E}_{\rm avg} \\
    \implies \dot{a}_{\rm avg} &=  - \frac{2 a^2}{G m_1 m_2} \dot{E}_{\rm avg}
    \label{eq:aavg}
\end{align}

Using Eqs. (\ref{eq:Pbarel}) and (\ref{eq:aavg}), we get,
\begin{align}
    \dot{P}_{\rm b,GR} = -\frac{3}{2} P_{\rm b}  \frac{2 a}{G m_1 m_2} \dot{E}_{\rm avg} 
    \label{eq:mlosspbdotgr}
\end{align}

For $\ddot{P}_{\rm b,GR}$, we take the time derivative of Eq. (\ref{eq:mlosspbdotgr}),

\begin{align}
    \ddot{P}_{\rm b,GR} &= -\frac{3}{2} \dot{P}_{\rm b,GR} \frac{2 a}{G m_1 m_2} \dot{E}_{\rm avg}  + \frac{3}{2} P_{\rm b} \left( \frac{4 a^2}{G^2 m_1^2 m_2^2} \dot{E}_{\rm avg}^2  - \frac{2 a}{G m_1 m_2} \ddot{E}_{\rm avg} \right)\nonumber \\
    &= -3 \dot{P}_{\rm b,GR} \frac{a}{G m_1 m_2} \dot{E}_{\rm avg}  + 3 P_{\rm b} \left( \frac{2 a^2}{G^2 m_1^2 m_2^2} \dot{E}_{\rm avg}^2  - \frac{a}{G m_1 m_2} \ddot{E}_{\rm avg} \right)
    \label{eq:pbddotgr0}
\end{align}

Using the expressions of $\dot{P}_{\rm b,GR}$ from Eq. (\ref{eq:Pbarel}) and $\dot{a}_{\rm avg}$ from Eq. (\ref{eq:aavg}) in Eq. (\ref{eq:pbddotgr0}), we get,

\begin{align}
\ddot{P}_{\rm b,GR} = 15 P_{\rm b} \frac{a^2}{G^2 m_1^2 m_2^2} \dot{E}_{\rm avg}^2 - 3 P_{\rm b}  \frac{a}{G m_1 m_2} \ddot{E}_{\rm avg} 
    \label{eq:pbddotgr1}
\end{align}

For calculating $\ddot{E}_{\rm avg}$, we first calculate the instantaneous value second derivative of $E$, that is, $\ddot{E}$ as follows,
\begin{align}
    &\ddot{E} = \frac{2G}{5c^5} \left[ \dddot{I}_{xx}\ddddot{I}_{xx} + \dddot{I}_{yy}\ddddot{I}_{yy} + 2 \dddot{I}_{xy}\ddddot{I}_{xy} \right. \nonumber \\
    &\left.\hskip2cm - \frac{1}{3}\left(\dddot{I}_{xx} + \dddot{I}_{yy}\right)\left(\ddddot{I}_{xx} + \ddddot{I}_{yy}\right)\right]~,\\
      \implies  &\ddot{E} = - \frac{8}{15c^5} \left(\frac{G^9 \mu^4 M^7}{a^{11} (1-e^2)^{11}}\right)^{1/2}  e\sin \theta \, (e \cos \theta +1)^3 \left(33 e^2 \cos \theta \right. \nonumber \\
    &\left.\hskip2cm +37 e^2+142 e \cos \theta +72\right)~.
    \label{eq:eddotI}
\end{align}

For calculating the average value, that is, $\ddot{E}_{\rm avg}$, 

\begin{align}
    \ddot{E}_{\rm avg} &= \frac{1}{P_{\rm b}} \int_{0}^{P_{\rm b}} \ddot{E} \,dt \nonumber \\ 
   &= \frac{1}{P_{\rm b}} \int_{0}^{2\pi} \ddot{E} \frac{dt}{d\theta} \,d\theta ~.
    \label{eq:Eddotavg}
\end{align}
 On substituting $\ddot{E}$ from Eq. (\ref{eq:eddotI}) and $\frac{d\theta}{dt}$ from Eq. (\ref{eq:thetadot}) in Eq. (\ref{eq:Eddotavg}), we get, $\ddot{E}_{\rm avg} = 0$.

Hence, Eq. (\ref{eq:pbddotgr1}) becomes,

\begin{align}
\ddot{P}_{\rm b,GR} = 15 P_{\rm b} \frac{a^2}{G^2 m_1^2 m_2^2} \dot{E}_{\rm avg}^2 \hskip0.5cm[\because \ddot{E}_{\rm avg}=0]
    \label{eq:nomlosspbddotgr}
\end{align}

\section{Standard Error Propagation for $\ddot{f}_{\rm b,GR}$}
\label{app:err}
We follow the standard error propagation technique to obtain the analytical expressions for the error terms corresponding to the parameters being calculated. The independent observables used to calculate $\ddot{f}_{\rm b,GR}$ are orbital period $P_{\rm b}$, eccentricity $e$, mass of companion $m_2$, and mass ratio $q_m = \frac{m_1}{m_2}$. Hence, the corresponding uncertainty expression for $\ddot{f}_{\rm b,GR}$ is given as,

\begin{align}
\Delta \ddot{f}_{\rm b,GR} &= \left( \left(\frac{\partial \ddot{f}_{\rm b,GR}}{\partial P_{\rm b}}\right)^2 {\Delta P_{\rm b}}^2 +  \left(\frac{\partial \ddot{f}_{\rm b,GR}}{\partial e}\right)^2 {\Delta e}^2 +  \left(\frac{\partial \ddot{f}_{\rm b,GR}}{\partial m_2}\right)^2 {\Delta m_2}^2 \right.\nonumber \\
&\left.+ \left(\frac{\partial \ddot{f}_{\rm b,GR}}{\partial q_m}\right)^2 {\Delta q_m}^2  \right)^{1/2}~.
    \label{eq:fb2err}
\end{align}

From Eq. (\ref{eq:nomlosspbddotgr0}) we have,

\begin{align}
\ddot{f}_{\rm b,GR} = \frac{3072}{25} \frac{G^{10/3}}{c^{10}} (2\pi)^{16/3} \frac{1}{P_{\rm b}^{19/3}} \frac{(1+\frac{73}{24}e^2+\frac{37}{96}e^4)^2}{(1-e^2)^{7}} \frac{q_m^{2} \,m_2^{10/3}}{(q_m + 1)^{2/3}}~.
    \label{eq:fbddotGrnomassloss}
\end{align}

Hence, the expressions for partial derivatives appearing in Eq. (\ref{eq:fb2err} are,

\begin{align}
  \frac{\partial \ddot{f}_{\rm b,GR}}{\partial P_{\rm b}} &=  -\frac{19456}{25} \frac{G^{10/3}}{c^{10}} \frac{(2\pi)^{16/3}}{P_{\rm b}^{22/3}} \frac{(1+\frac{73}{24}e^2+\frac{37}{96}e^4)^2}{(1-e^2)^{7}} \frac{q_m^{2} \,m_2^{10/3}}{(q_m + 1)^{2/3}} \\
    \frac{\partial \ddot{f}_{\rm b,GR}}{\partial e} &=  \frac{6144}{25} \frac{G^{10/3}}{c^{10}}  \frac{(2\pi)^{16/3}}{P_{\rm b}^{19/3}} \frac{(\frac{157}{12}e+\frac{18517}{288}e^3+\frac{92989}{1152}e^5 +\frac{9953}{768}e^7 +\frac{1369}{3072}e^9)}{(1-e^2)^{8}} \nonumber \\
    &\hskip1cm \times \frac{q_m^{2} \,m_2^{10/3}}{(q_m + 1)^{2/3}} \\
 \frac{\partial \ddot{f}_{\rm b,GR}}{\partial m_2} &= \frac{10240}{25} \frac{G^{10/3}}{c^{10}}  \frac{(2\pi)^{16/3}}{P_{\rm b}^{19/3}} \frac{(1+\frac{73}{24}e^2+\frac{37}{96}e^4)^2}{(1-e^2)^{7}} \frac{q_m^{2} \,m_2^{7/3}}{(q_m + 1)^{2/3}} \\
     \frac{\partial \ddot{f}_{\rm b,GR}}{\partial q_m} &= \frac{2048}{25} \frac{G^{10/3}}{c^{10}} \frac{(2\pi)^{16/3}}{P_{\rm b}^{19/3}} \frac{(1+\frac{73}{24}e^2+\frac{37}{96}e^4)^2}{(1-e^2)^{7}} \frac{m_2^{10/3}\,(2q_m^2 + 3q_m)}{(q_m + 1)^{5/3}} 
\end{align}

Table (\ref{tb:fddotgrrealerr}) shows $\ddot{f}_{\rm b,GR}$ values as well as corresponding errors calculated using the standard error propagation technique for five real pulsars. For these pulsars (PSRs J0348+0432, J0740+6620, J1738+0333, J2045+3633, and J2339-0533), the values as well as errors of $P_{\rm b}$, $e$,  $m_2$, $q_m = m_1/m_2$ are available in ATNF pulsar catalogue \citep{mhth05}.

\begin{table*}[h]
\begin{center}
\caption{Parameters of a set of real pulsars with corresponding errors calculated using standard error propagation technique. The errors are shown in parenthesis next to the decimal place corresponding to their own exponent. Here, $P_{\rm b}$ is the orbital period, $e$ is the eccentricity,  $m_2$ is the companion mass, $q_m = m_1/m_2$ is the mass ratio, and $\ddot{f}_{\rm b,GR}$ is the general relativistic contribution to the orbital frequency second derivative.}
\hspace*{-1cm}\begin{tabular}{|@{\hskip1.pt}l@{\hskip1.pt}|@{\hskip1.pt}l@{\hskip1.pt}|@{\hskip1.pt}l@{\hskip1.pt}|@{\hskip1.pt}l@{\hskip1.pt}|@{\hskip1.pt}l@{\hskip1.pt}|@{\hskip1.pt}l@{\hskip1.pt}|}
 \hline 
     \diagbox{Parameters}{PSR} & J0348+0432 & J0740+6620 & J1738+0333 & J2045+3633 & J2339-0533  \\  
\hline
\addstackgap{$P_{\rm b}$ (days)}  & 0.102424063(3) &  4.76694462(4)  &  0.3547907399(5)  &  32.29784(3) &  0.19309840(1)  \\   
\hline
\addstackgap{$e$} & 2(1)$\times10^{-6}$ & 6.01(4)$\times10^{-6}$  &  3(1)$\times10^{-7}$  &  0.017212447(6)  &  0.0002102(5)  \\     
\hline   
\addstackgap{$m_2$ (kg)} & 3.42(6)$\times10^{29}$ &   5.1(2)$\times10^{29}$  &  3.6(1)$\times10^{29}$  &  1.74(3)$\times10^{30}$  &  7(1)$\times10^{29}$ \\   
\hline 
\addstackgap{$q_m$} & 11.7(1) &   8.22(8) &  8.1(2) &  1.43(4)  &  4.6(2) \\  
\hline 
\addstackgap{$\ddot{f}_{\rm b,GR}$ ($\rm s^{-3}$)}& 3.2(2)$\times10^{-38}$ &    2.1(2)$\times10^{-48}$  &  9(1)$\times10^{-42}$  &  4.9(3)$\times10^{-53}$  &  1.68(96)$\times10^{-39}$ \\ 
\hline
\end{tabular}
\label{tb:fddotgrrealerr}
\end{center}
\end{table*}
 \clearpage

\nocite{*}

\bibliography{PathakPRD}

\end{document}